# A comprehensive study on the relationship between image quality and imaging dose in low-dose cone beam CT

**Hao Yan, Laura Cervino, Xun Jia, and Steve B. Jiang**

Center for Advanced Radiotherapy Technologies and Department of Radiation Medicine and Applied Sciences, University of California San Diego, La Jolla, CA 92037-0843, USA

E-mails: sbjiang@ucsd.edu, xunjia@ucsd.edu

While compressed sensing (CS) based algorithms have been developed for low-dose cone beam CT (CBCT) reconstruction, a clear understanding on the relationship between the image quality and imaging dose at low dose levels is needed. In this paper, we qualitatively investigate this subject in a comprehensive manner with extensive experimental and simulation studies. The basic idea is to plot both the image quality and imaging dose together as functions of number of projections and mAs per projection over the whole clinically relevant range. On this basis, a clear understanding on the tradeoff between image quality and imaging dose can be achieved and optimal low-dose CBCT scan protocols can be developed to maximize the dose reduction while minimizing the image quality loss for various imaging tasks in image guided radiation therapy (IGRT). Main findings of this work include: 1) Under the CS-based reconstruction framework, image quality has little degradation over a large range of dose variation. Image quality degradation becomes evident when the imaging dose (approximated with the x-ray tube load) is decreased below 100 total mAs. An imaging dose lower than 40 total mAs leads to a dramatic image degradation, and thus should be used cautiously. Optimal low-dose CBCT scan protocols likely fall in the dose range of 40-100 total mAs, depending on the specific IGRT applications. 2) Among different scan protocols at a constant low-dose level, the super sparse-view reconstruction with projection number less than 50 is the most challenging case, even with strong regularization. Better image quality can be acquired with low mAs protocols. 3) The optimal scan protocol is the combination of a medium number of projections and a medium level of mAs/view. This is more evident when the dose is around 72.8 total mAs or below and when the ROI is a low-contrast or high-resolution object. Based on our results, the optimal number of projections is around 90 to 120. 4) The clinically acceptable lowest imaging dose level is task dependent. In our study, 72.8mAs is a safe dose level for visualizing low-contrast objects, while 12.2 total mAs is sufficient for detecting high-contrast objects of diameter greater than 3 mm.

**1. Introduction**

Flat panel x-ray cone beam CT (CBCT) imaging has become an important tool in image guided radiotherapy (IGRT) (Jaffray *et al.*, 1999; Jaffray *et al.*, 2002; McBain *et al.*, 2006; Grills *et al.*, 2008). The daily use of CBCT imaging produces a considerable amount of excessive radiation dose to radiotherapy patients (Islam *et al.*, 2006; Daly *et al.*, 2006; Ding and Coffey, 2009), making it necessary to lower the imaging dose especially for young patients (Kim *et al.*, 2010). Generally speaking, low-dose scanning can be achieved by either fixing the number of projections while decreasing the x-ray tube load (mAs) or fixing the mAs level while decreasing the number of projections. When the CBCT images are reconstructed using the classic Feldkamp-Davis-Kress (FDK) algorithm (Feldkamp *et al.*, 1984), the low-mAs scheme results in noisy reconstructed images due to the low signal-to-noise ratio (SNR) in the projection images, while the sparse-view scheme leads to severe under-sampling streaking artifacts. Recent breakthrough in compressed sensing (CS) theory (Candes *et al.*, 2006; Candes and Tao, 2006; Donoho, 2006) has greatly stimulated research efforts on biomedical imaging (Wang *et al.*, 2011), especially low-dose CT/CBCT reconstruction. Under the CS-based reconstruction framework, a priori sparsity property of the images could be utilized in iterative CT/CBCT reconstructions through Total Variation (TV) minimization (Sidky *et al.*, 2006; Song *et al.*, 2007; Sidky and Pan, 2008; Choi *et al.*, 2010; Jia *et al.*, 2010; Defrise *et al.*, 2011; Ritschl *et al.*, 2011; Tian *et al.*, 2011), soft-thresholding filtering (Yu and Wang, 2010), Tight Frame (TF) regularization (Jia *et al.*, 2011), or by using the prior images (Chen *et al.*, 2008; Leng *et al.*, 2008; Xu *et al.*, 2011), which dramatically reduces the information required for reconstruction. Compared with the classic FDK algorithm, CS-based reconstruction algorithms are able to handle both sparse-view reconstruction and low-mAs reconstruction, achieving a comparable image quality at a much lower imaging dose level.

However, the image quality will be degraded inevitably when the imaging dose decreases. It is necessary to explore the relationship between image quality and imaging dose, which may provide guidance for designing optimal low-dose scan protocols for various specific IGRT tasks. Some preliminary work has been done along this direction. Han *et al.* have compared three different dose allocation schemes at a constant dose level, with a focus on arguing that their CS-based method outperforms the FDK algorithm (Han *et al.*, 2010). While comparing the CS-based and the classic statistical-based iterative reconstruction methods, Tang *et al.* found that it is preferable to distribute total imaging dose into many view angles to reduce the under-sampling streaking artifacts (Tang *et al.*, 2009). However, these studies are both based on a single dose level, and more comprehensive work is still needed.

In this paper, we present a comprehensive study on the relationship between image quality and imaging dose under the CS reconstruction framework using an in-house developed TF-based algorithm (Jia *et al.*, 2011). This algorithm has been found to have similar performance as (Jia *et al.*, 2011) and a theoretical link to (Cai *et al.*, 2011) the more widely used TV minimization method. Our basic idea is to generate a dose-quality map (DQM) by plotting both the image quality and imaging dose together as functions of number of projections and mAs per projection (mAs/view) over the whole clinically relevant range. With the DQM, we can have a clear global picture of the tradeoff between image quality and imaging dose and, thus, design optimal low-dose CBCT scan protocols that maximize dose reduction while minimizing image quality loss for various IGRT imaging tasks. On the first order approximation, imaging dose is proportional to tube load (mAs). In this paper, we therefore use tube load to approximate imaging dose for the sake of simplicity.

## 2. Methods and Materials

*2.1 Data acquisition*

*2.1.1 CBCT system and experimental phantom*

A kV on-board imaging (OBI) system integrated in a TrueBeam$^{TM}$ medical linear accelerator (Varian Medical System, Palo Alto, CA) was used in the experiment. The x-ray tube has a small focal spot of 0.4 mm and a 2.7 mm inherent filtration on the exit window. The imaging detector has 768×1024 pixels and the pixel pitch in each dimension is 0.392 mm, which yields an active detector area of 30.1×40.1 cm$^2$. An anti-scatter grid is mounted on the detector. The distances from the source to the rotation center and to the detector plane are 100 cm and 150 cm, respectively. A CatPhan® 600 phantom (The Phantom Laboratory, Inc., Salem, NY) was scanned in a full-fan mode with a full-fan bow-tie filter on site (figure 1a). In each scan, 364 projections over 200 degree were acquired.

*2.1.2 Measured data points*

The tube voltage was fixed to 100kVp in all scans. The CatPhan® 600 phantom was scanned under thirteen dose levels, *i.e.*, 73.2, 95.6, 109.8, 146.4, 161.04, 190.32, 219.6, 292.8, 329.4, 395.3, 439.2, 585.6 and 878.4 total mAs, corresponding to 0.2, 0.26, 0.3, 0.4, 0.44, 0.52, 0.6, 0.8, 0.9, 1.08, 1.2, 1.6 and 2.4 mAs/view. For each dose level, 46, 61, 91, 121, 182, and 364 projections were extracted with equal angular intervals from the total 364 projections, so that the final experimental database consists of 13×6 data points on the DQM, including four groups of data points of equal dose values around 36.4, 72.6, 109.2 and 146.4 total mAs (figure 1b).

*2.1.3 Measurement-based simulation*

Although a large database was collected, the measured data is still not dense enough to cover the whole DQM over the clinically relevant range. The main reason is because the minimum angle between two neighboring projections in the OBI system is fixed to 200/360·2π/364 (full-fan mode), and consequently one cannot get evenly distributed projections with total projection numbers between 182 and 364 (figure 1b). To have a complete view of DQM, we perform a measurement-based simulation as a supplement to the experiment.

A digital phantom (figure 1c) was specifically designed for this simulation according to the contrast module of CatPhan® 600 phantom. The forward x-ray projections were calculated using an analytical approach based on Tang *et al* (2006). To be more realistic, a non-uniform fluence map caused by the bow-tie filter was measured and used in this simulation. The simulated scanning configuration was exactly the same as that in the experiment. A 12×12 data grid was generated (figure 1d). The projection number varies from 30 to 360 at an interval of 30. To simulate different mAs levels, we superimposed noises based on the measured data. The noise model employed here was proposed by (Li *et al.*, 2004) based on experiment data and has been further validated in (Wang *et al.*, 2008). Specifically, the standard deviation of a pixel value can be described by the following formula:

$$std(g_i) = b_i \exp\left(\frac{\mu(g_i)}{T}\right) \quad (1)$$

where $\mu$ represents the mean, $g_i$ is the detected signal at the pixel $i$, $T$ is a machine-dependent constant, and $b_i$ is determined by the mAs/view and the fluence map. In commissioning the noise model against our experiment, the axis-invariant sinogram data scanned with 2.4mAs/view were averaged to estimate $\mu(g_i)$. One can get the variance for each data point at different mAs levels, and then $T$ and $b_i$ can be obtained according to the steps in (Li *et al.*, 2004).

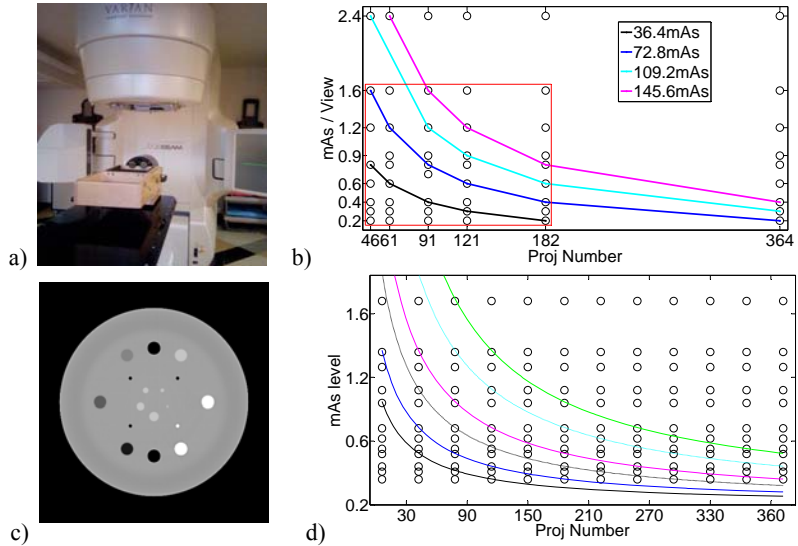

**Figure 1.** Data acquisition equipment and scheme. a) Catphan® 600 phantom and the CBCT scanner used in the experiment; b) The measurement data grid points and four iso-dose lines around 36.4, 72.8, 109.2 and 145.6 total mAs. Only the dense grid points inside the red box will be used to produce the DQM; c) The digital phantom used in the simulation with display grayscale in terms of attenuation coefficient: [0.018, 0.025] mm$^{-1}$; d) The simulation data grid points and six iso-dose lines of 24.3, 36.4, 54.6, 72.8, 109.2 and 145.6 total mAs

*2.2 CBCT reconstruction algorithm*

In this study, we use a recently developed CBCT reconstruction algorithm with TF regularization (Jia *et al.*, 2011). A detailed description of the TF reconstruction algorithm can be found in (Jia *et al.*, 2011).

Iterative volume reconstruction itself is computationally extensive due to the multiple operations of back-projection and forward-projection. Therefore the TF algorithm has been implemented on the platform of computer graphics processing unit (GPU) (Jia *et al.*, 2011). Moreover, in this study reconstructions have to be performed for all 13×6 data points and for each data point, various values of the regularization parameter $\beta$ (μ in (Jia *et al.*, 2011)) have to be tested to obtain the optimal reconstruction. For instance, to test 10 $\beta$ values, a total number of 780 (13×6×10) reconstructions of images of size 512×512×140 are needed, which is a huge computational task. We then implemented the GPU-based TF code in a batch mode.

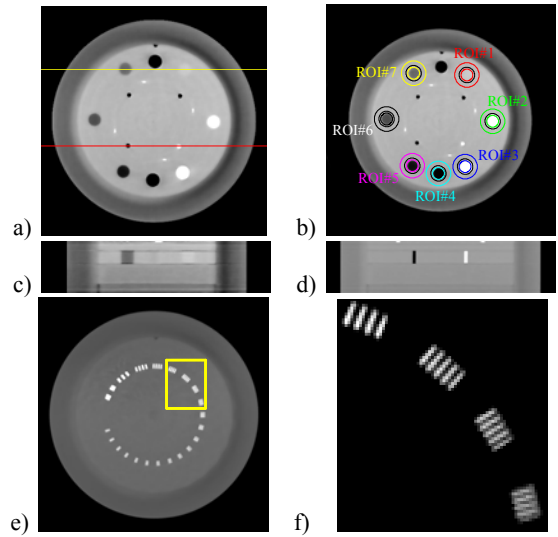

**Figure 2.** Modules in the Catphan® 600 phantom used in this study. a) A contrast slice shown with [0.018, 0.025] mm$^{-1}$; b) Seven ROIs in the contrast slice; c) The slices containing low contrast objects as indicated by the yellow line in a), shown with [0.018, 0.025] mm$^{-1}$; d) The slices containing small high contrast objects as indicated by the red line in a), shown with [0.008, 0.032] mm$^{-1}$; e) A resolution slice shown with [0.008, 0.045]; f) The ROI in e) (yellow box) shown with [0.03, 0.05] mm$^{-1}$.

### 2.3. Image quality assessment (IQA)

It has been reported that high-contrast targets can be well reconstructed by the CS-based reconstruction from super sparse views (Jia *et al.*, 2010), which is sufficient for bony structure based patient positioning in IGRT. A more challenging task in IGRT is soft tissue based patient positioning, which is the main reason for the wide adoption of CBCT in IGRT. In this study we have put more focus on the visualization and detection of low contrast objects.

### 2.3.1 Visual inspection

To reveal the contrast details and catch the artifacts that may otherwise be invisible in a wide window, we observed the reconstructed images containing low-contrast objects (figure 1c, figure 2a-c) with a consistent narrow grayscale window of [0.018, 0.025] mm$^{-1}$, which corresponds to a soft-tissue display window of 375 HU width. The difference between each image and its reference image was also visually investigated with a grayscale window of [-0.00125, 0.00125] mm$^{-1}$. We used a wide grayscale window of [0.008, 0.032] mm$^{-1}$ for the image containing high-contrast air and Teflon robs with 3mm diameter (figure 2d). For the resolution slice (figure 2e) and the region of interest (ROI) therein (figure 2f), [0.008, 0.045] mm$^{-1}$ and [0.03, 0.05] mm$^{-1}$ were used, respectively.

*2.3.2 IQA metrics*

To generate quantitative IQA scores, we used a number of IQA metrics, including a classic difference-based metric, *i.e.*, relative root mean square error (rRMSE), and a recently proposed perception-based metric, *i.e.* feature similarity index (FSIM). For the experimental data, contrast-noise ratio (CNR) was also used to further investigate the visibility of the ROIs shown in figure 2b.

*2.3.2.1. rRMSE*   The conventional rRMSE is defined as

$$\text{rRMSE}_1 = \sqrt{\frac{\sum_{i,j}|f_{i,j} - f_{i,j}^{\text{ref}}|^2}{\sum_{i,j}|f_{i,j}^{\text{ref}}|^2}} \quad (2)$$

where $f$ and $f^{\text{ref}}$ represent the evaluation and reference image, respectively, and $i, j$ labels the pixel coordinates in the 2D images. To focus more on the low contrast objects, we define a second rRMSE as

$$\text{rRMSE}_2 = \sqrt{\sum_{i,j}\left(\frac{|f_{i,j} - f_{i,j}^{\text{ref}}|}{w_{i,j}}\right)^2} \quad (3)$$

with

$$w_{i,j} = \exp\left(\left|\frac{f_{i,j}^{\text{ref}} - \mu_{\text{water}}}{\mu_{\text{water}} - \mu_{\text{polystyrene}}}\right|\right) \quad (4)$$

Compared with *rRMSE*$_1$, *rRMSE*$_2$ is more sensitive to the bias introduced by low-contrast objects with attenuation coefficients around $\mu_{water}$. In this work, $\mu_{water}$ and $\mu_{polystyrene}$ were set to 0.0206 *mm*$^{-1}$ and 0.0196 mm$^{-1}$, respectively.

*2.3.2.2. FSIM*   It has been recognized that IQA could be more consistent with the human subjective evaluation if the metrics are designed by considering the mechanism of the human visual system. Perception-based metrics were proposed for this purpose. The basic idea is to perform IQA on the structure level rather than the pixel level, mimicking how human visual system works. A novel metric called feature similarity (FSIM) index was recently proposed (Zhang *et al.*, 2011), where the main features are the phase congruency (PC) of the local structure and the image gradient magnitude (GM). Extensive experiments on six benchmark IQA databases demonstrated that FSIM could achieve higher consistency with the subjective evaluation than other state-of-the-art IQA metrics

(Zhang *et al.*, 2011). We therefore have used FSIM as an IQA metric in this work. FSIM is defined as

$$\text{FSIM} = \frac{\sum_{i,j}\{S_{PC} \cdot S_{GM} \cdot PC_m\}}{\sum_{i,j}\{PC_m\}} \quad (5)$$

by using both PC-based similarity metrics $S_{PC}$

$$S_{PC} = \frac{PC(f_{i,j}^{\text{ref}}) \cdot PC(f_{i,j}) + T_1}{[PC(f_{i,j}^{\text{ref}})]^2 + [PC(f_{i,j})]^2 + T_1} \quad (6)$$

and GM-based similarity metrics $S_{GM}$

$$S_{GM} = \frac{GM(f_{i,j}^{\text{ref}}) \cdot GM(f_{i,j}) + T_2}{[GM(f_{i,j}^{\text{ref}})]^2 + [GM(f_{i,j})]^2 + T_2} \quad (7)$$

with a PC-based weighting factor $PC_m$

$$PC_m = max[PC(f_{i,j}^{\text{ref}}), PC(f_{i,j})]. \quad (8)$$

In (6) and (7), $T_1$ and $T_2$ are positive constants to increase the calculation stability. More detailed descriptions of PC and GM are given in (Zhang *et al.*, 2011). We directly used the code[1] provided by the authors in our evaluation. The FSIM score varies between 0 and 1 with 1 representing the best image quality.

*2.3.2.3. CNR*   CNR was used for the ROI visibility evaluation for the experimental data. As seen in the contrast slice (figure 2b), seven areas within the inner circle were selected as the ROI targets and the corresponding concentric rings were used as backgrounds. The CNR is defined as

$$\text{CNR} = \frac{|\mu(f_{\text{ROI}}^t) - \mu(f_{\text{ROI}}^b)|}{std(f_{\text{ROI}}^t) + std(f_{\text{ROI}}^b)} \quad (9)$$

where the superscripts *t* and *b* represent the target and background, respectively.

*2.4 Dose-quality relationship evaluation*

*2.4.1 DQM generation*

DQMs were generated based on the simulated data (figure 1c) and measured data (figures 2a and 2c). Compared with the subjective visual inspection methods for tuning β as in many other works, in this work we combine a subjective inspection method with objective metrics to take advantage of each method's strengths. Human visual inspection is commonly considered as the gold standard for image quality assessment. Currently, no IQA metric performs equally as human visual system. Each metric is more or less biased by emphasizing certain aspects of the image quality. However, it is hard to determine the best image from several very similar candidates with visual inspection. In that case, objective metrics can be used to help identify the best image with a quantitative score.

---

[1] http://www4.comp.polyu.edu.hk/~cslzhang/IQA/FSIM/FSIM.htm

The detailed steps are as follows,
1) A reference image was reconstructed as the gold standard for evaluation. For the simulation data, the reference image was reconstructed from 1080 noise-free projections, while for the experimental data, the reference image was reconstructed from 364 projections scanned at the highest imaging dose (2.4 mAs/view and 873.6 total mAs).
2) Three $β$ values with similarly good visual performance were selected manually as candidates.
3) IQA metrics were then used to determine the optimal $β$ value. If for a data point different optimal $β$ values were indicated by different metrics, they were averaged to obtain the final optimal $β$ value.
4) The DQM was generated by interpolating the image quality scores of all data points as shown in figures 1b and d. Note that due to the accuracy consideration, for the experimental data, only the dense data points inside the red box in figure 1b were used.

*2.4.2 Further evaluation*

Further analysis was performed for data points on the iso-dose lines in DQM's. For the images containing low-contrast targets (figure 2a and c), we closely inspected them using visual inspection and rRMSE, FSIM, and CNR. We also inspected the resolution slice (figure 2f) and the slices containing high-contrast targets (figure 2a and d).

**3. Results**

*3.1 Simulation data*

DQMs in terms of $rRMSE_1$, $rRMSE_2$, and FSIM are shown in figures 3a-c, with five iso-dose lines superimposed. For these three figures, the x-axis is the number of projections ranging from 30 to 360, y-axis is the mAs/view ranging from 0.2 to 1.6, and the color represents the IQA scores ($rRMSE_1$, $rRMSE_2$, and FSIM for figures 3a, 3b, 3c, respectively). The six iso-dose lines correspond to total mAs levels of 24.3, 36.4, 54.6, 72.8, 109.2, and 145.6. In figure 3c, the white dotted line with arrowhead indicates the steepest image quality variation direction. Along this direction, the variation of the IQA scores with total mAs is plotted in figure 3e. The optimal $β$ values for different data points are illustrated by different colors in figure 3d.

From figures 3a-c, we can observe that
1) Image quality has little degradation over a large range of imaging dose variation (upper-right corner) but changes rapidly at the low dose range (lower-left corner).
2) For cases with extremely sparse views (projection number < 50), image quality degrades severely no matter how high the mAs/view is.
3) For cases with very low mAs/view (*e.g.*, 0.4 mAs/view), reasonable image quality may still be achieved with a large amount of views (*e.g.*, 360 projections).
4) The iso-dose lines are hyperbolic functions while in low-dose range; the iso-quality lines have larger curvature than hyperbolic functions. Therefore, there exists a point of maximum image quality on an iso-dose line or a point of minimum imaging dose on an iso-quality line.

5) The white dotted line with arrowhead in figure 3c represents the steepest image quality variation direction. The points of maximum image quality on iso-dose lines are near the intersection points of the iso-dose lines and the white dot line, indicating that the optimal scan protocol in low-dose range consists of a medium number of projections and a medium mAs/view level.

Two more observations can be drawn from figure 3d and 3e:

6) Larger $\beta$ values are needed with the lower imaging dose levels, and specifically, very large $\beta$ values are needed for the sparse-view case (figure 3d).

7) Image quality varies slowly over a large imaging dose range above 100 total mAs, more quickly when below 100 total mAs, and at a very large rate when below 40 total mAs (indicated by the pink dashed line in figure 3e).

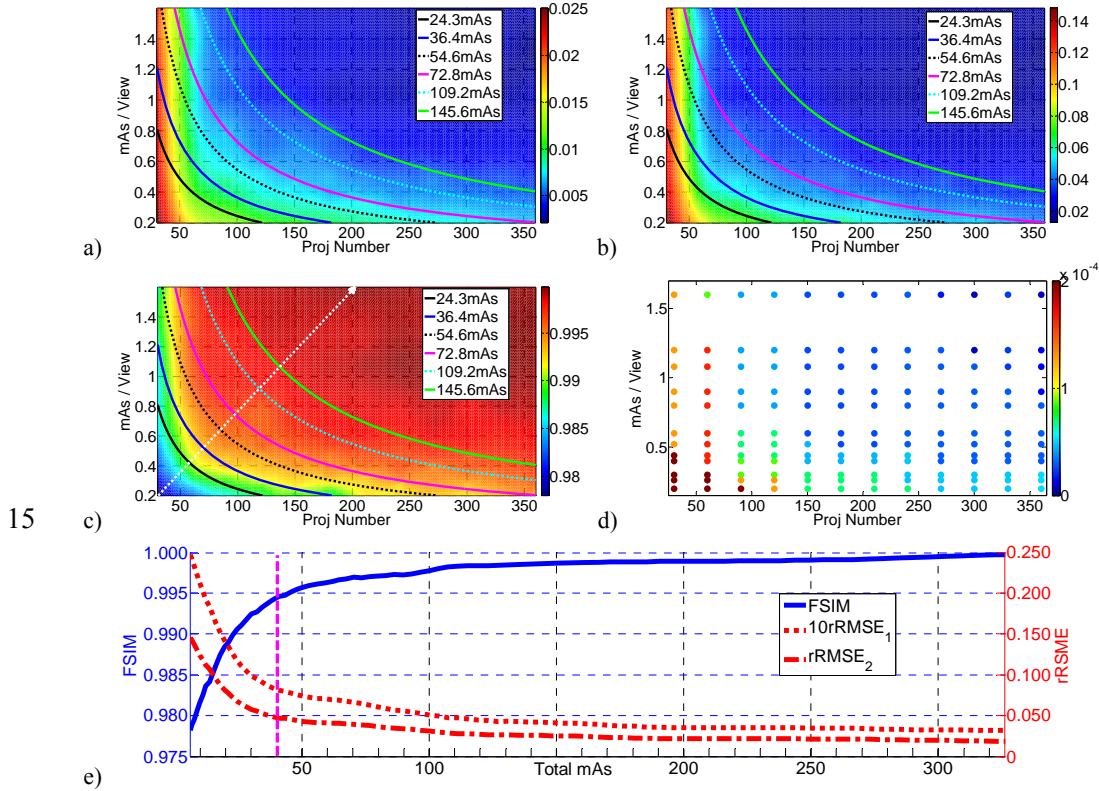

**Figure 3.** Results of the simulation data. DQMs for a) $rRMSE_1$; b) $rRMSE_2$; c) FSIM with 6 iso-dose lines at total mAs levels of 24.3, 36.4, 54.6, 72.8, 109.2, and 145.6. The white dot line with arrow head in c) indicates the steepest image quality variation direction. d) The optimal $\beta$ values for different data points. e) The variation of the IQA scores with the total mAs, along the white dot line in c). The pink dashed vertical line indicates 40 total mAs below which the IQA scores degrade at a steeper rate.

*3.2 Measurement data*

*3.2.1 DQM*

Similar to figure 3, the results for the measurement data are shown in figure 4. The DQMs in terms of $rRMSE_1$, $rRMSE_2$ and FSIM are shown in figure 4a, with data points representing four iso-dose levels around 36.4, 72.8, 109.2, and 145.6 total mAs. Only the dense data points (figure 1b) were used for building these DQMs to ensure accuracy. In figure 4b, the IQA scores of all data points are plotted with logarithm-scale applied to the horizontal axis. The optimal $\beta$ values for different data points are shown in figure 4c.

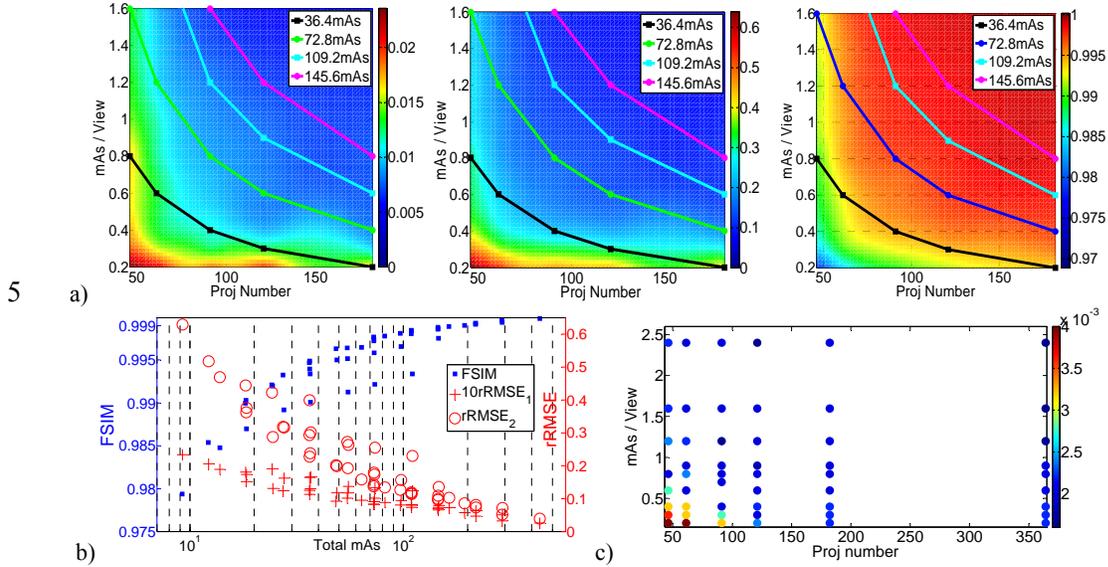

**Figure 4.** Results of the measurement data. a) From left to right: DQMs for $rRMSE_1$, $rRMSE_2$, and FSIM with data points at four iso-dose lines around 36.4, 72.8, 109.2, and 145.6 total mAs. b) The variation of the IQA scores with dose decreasing from 576 to 6 total mAs; c) The optimal β values for different data points.

Basically, figure 4 verifies figure 3; the observations from figure 4 are consistent with those from figure 3. From figure 4a, it is clear that the optimal image quality is obtained with a protocol of medium number of views and medium mAs/view. As a discrete counterpart of figure 3e, figure 4b also exhibits that image quality degradation is not evident with imaging doses from 576 to 100 total mAs, and at a much steeper rate after 40 total mAs. Moreover, the IQA scores vary at the same total mAs value, indicating the image quality varies even for the same dose level due to the use of different scan protocols. Figure 4c illustrates that larger $\beta$ values are needed for low dose levels, consistent with figure 3d.

*3.2.2 Low-contrast objects: image quality on the iso-dose line*

*3.2.2.1. Visual inspection*   Corresponding to the four iso-dose levels (36.4, 72.8, 109.2 and 145.6 total mAs), the images with different scan protocols are shown in figures 5-8, in which our focus is on a low-contrast object indicated by the yellow arrow in the reference image (last column, figure 5). A specific protocol is represented by *a×b(c)*, where *a* represents the number of projections, *b* is the mAs/view, and *c* is the actual total mAs. Note that some data points are not available due to the experiment limitation on the fixed minimum view interval and discretely collected mAs levels. For example, 364×0.1 is not available for the 36.4 total mAs level (figure 5) because 0.2mAs is the minimum mAs provided by the machine. For a similar reason, tiny differences existed within the iso-dose level, *e.g.* the actual values for the 36.4 total mAs level are between 36.3 and 36.8, which are labeled accordingly in figure 5-8.

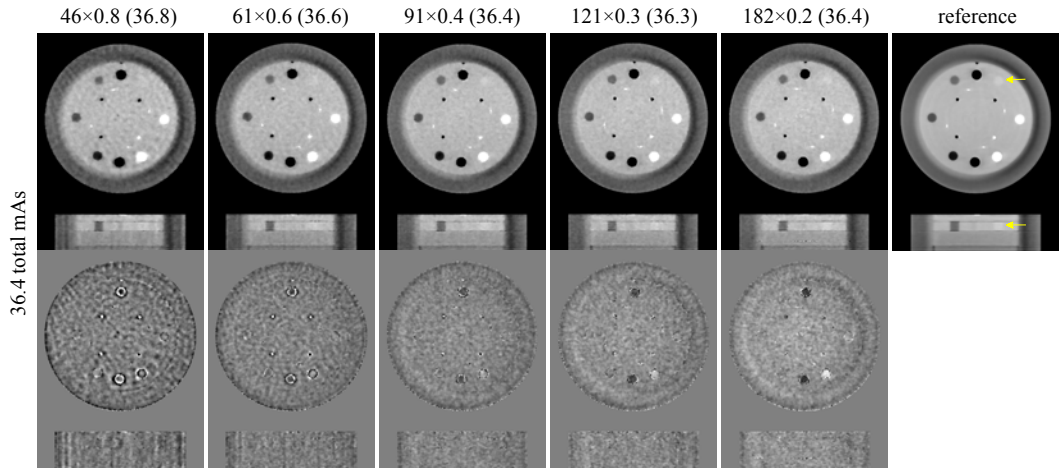

**Figure 5.** Images from different scan protocols with total dose around 36.4 mAs. Upper row: reconstructed images; display grayscale: [0.018, 0.025] mm$^{-1}$. Bottom row: difference between the reconstructed images and the reference image; display grayscale: [-0.00125, 0.00125] mm$^{-1}$. The low-contrast object at one o'clock direction is indicated by yellow arrows in the reference image. For each protocol $a\times b(c)$, $a$ represents the number of projections, $b$ is the mAs/view, and $c$ is the actual total mAs.

From figure 5, it is clear that severe artifacts exist in the sparse-view image (46×0.8), where the low contrast object is indistinguishable. The artifacts are much alleviated (yet still visible) in 61×0.6, and the low contrast object starts emerging. It can be seen that the five reconstructed images can be ranked from high to low quality based on the visibility of the low-contrast object and the difference images in the following order: 91×0.4, 121×0.3, 182×0.2, 61×0.6, and 46×0.8, which is consistent with our observations from figures 3 and 4.

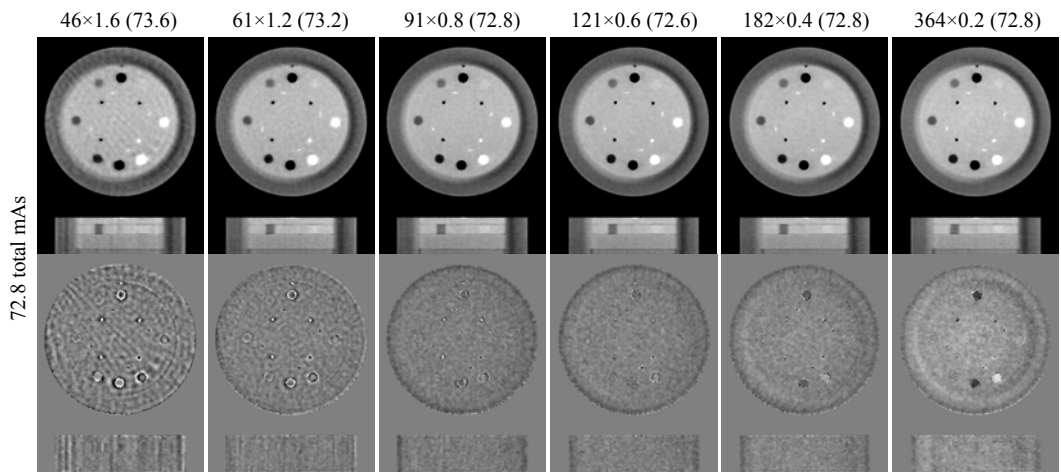

**Figure 6.** Same as figure 5 except that the total mAs is around 72.8.

The observations based on the difference images form figures 6-8 are consistent with that from figure 5, *i.e.*, scan protocols with a medium number of projections (~91-121) and a medium level of mAs/view yield reconstruction images of optimal quality. However, based on the visibility of the low-contrast object, the quality of the images scanned with the number of projections greater than 91 is very similar, which is also consistent with figures 3 and 4. We also noticed that it is evident that high contrast objects degrade dramatically in the image reconstructed from more projections. When the imaging dose level is fixed, using more projections means a lower SNR in each projection. In such cases, image degradation is expected and all objects will be degraded to some extent. However, it is more profound for high-contrast object likely due to the large intensity difference. Hence, when plotting the difference image at a certain window level, the degradation of high-contrast object is more obvious.

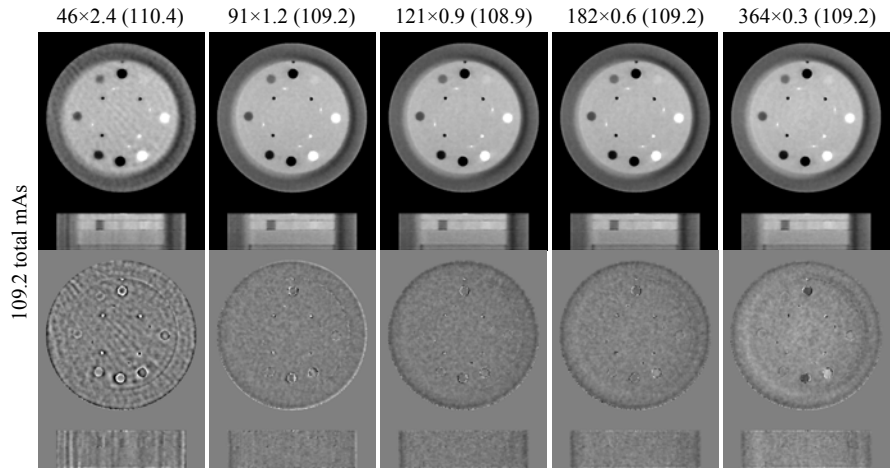

**Figure 7.** Same as figure 5 except that the total mAs is around 109.2.

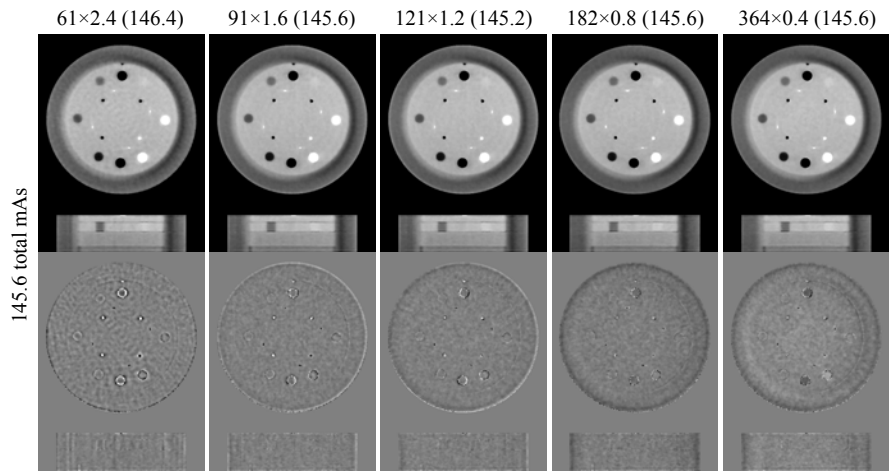

**Figure 8.** Same as figure 5 except that the total mAs is around 145.6.

*3.2.2.2. IQA scores* For the images in figures 5-8, the corresponding $rRMSE_1$, $rRMSE_2$ and FSIM scores, as well as the CNR scores for different ROIs are shown in figures 9 and 10.

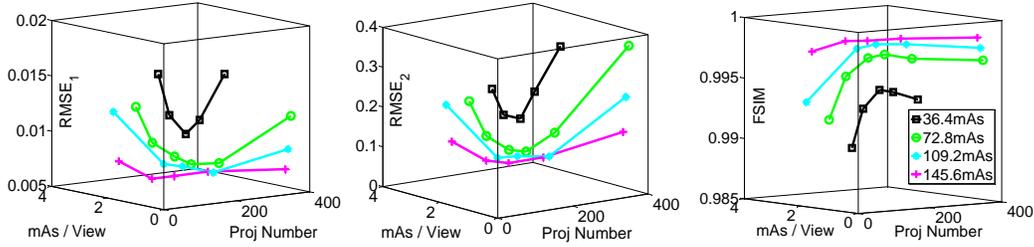

**Figure 9.** IQA scores for the images containing low-contrast objects at different iso-dose levels.

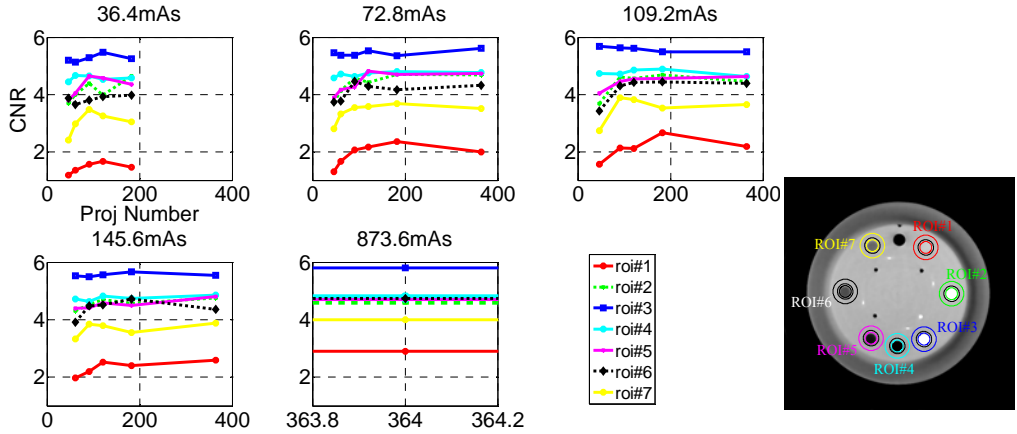

**Figure 10.** CNR scores for the seven ROIs from low- to high-contrast objects at four iso dose levels.

In figure 9, all the global metrics show the same phenomenon as observed from the visual inspection (figures 5-8), *i.e.*, the optimal image quality can be achieved at the middle of each iso-dose line, and this phenomenon is more evident at the lower dose levels such as 36.4 and 72.8 total mAs. We can also see that image quality has a considerable improvement when the dose is increased from 36.4 to 72.8 total mAs, while less considerable for the total mAs changes from 72.8 to 145.6.

In figure 10, the CNR scores are given for different total dose levels and different ROIs. We also plot the CNR scores at 873.6 total mAs as a reference. From this figure, we can see that the maximum CNR scores for each ROI locate in the range of 91-180 projections, which again is consistent with the observations from figures 3-9. This phenomenon is more obvious for the two low-contrast ROIs at the one and eleven o'clock directions (yellow and red) and at lower dose levels. For the high-contrast ROIs such as Teflon and air at the three and six o'clock directions (blue and cyan), this phenomenon is not obvious. Furthermore, the CNR only decreases for the low-contrast ROIs (black, yellow, and red) and does not change significantly for high-contrast ROIs. Specifically, if we look at the red ROI at the one o'clock direction, which is the low-contrast object used for visual detection in figure 5-8, we can find that its CNR changes most significantly with the dose variation.

*3.2.3 High-resolution objects: image quality on the iso-dose line*

The reconstructed ROI images in the resolution slice are shown in figure 11. In the reference image, we can discriminate up to 9 line pairs per cm (the bottom-right corner of the image panel). For the 36.4 total mAs level, we can only roughly visualize 8 line pairs. For the 72.8 total mAs level, we can identify 8 line pairs per cm clearly with all the scan protocols, and almost 9 line pairs per cm with protocol of 121×0.6. When the dose is increased to 109.2 and 145.6 total mAs, the overall clarity is improved.

The general trend observed from the high-resolution cases is quite similar to that from the low-contrast cases. It seems that the optimal dose allocation scheme of 72.8 total mAs, *i.e.* 121×0.6 offers an acceptable resolution compared with the reference (873.6 total mAs). If one needs a more robust and easy discrimination of 9 line pairs per cm, over 100 total mAs is necessary.

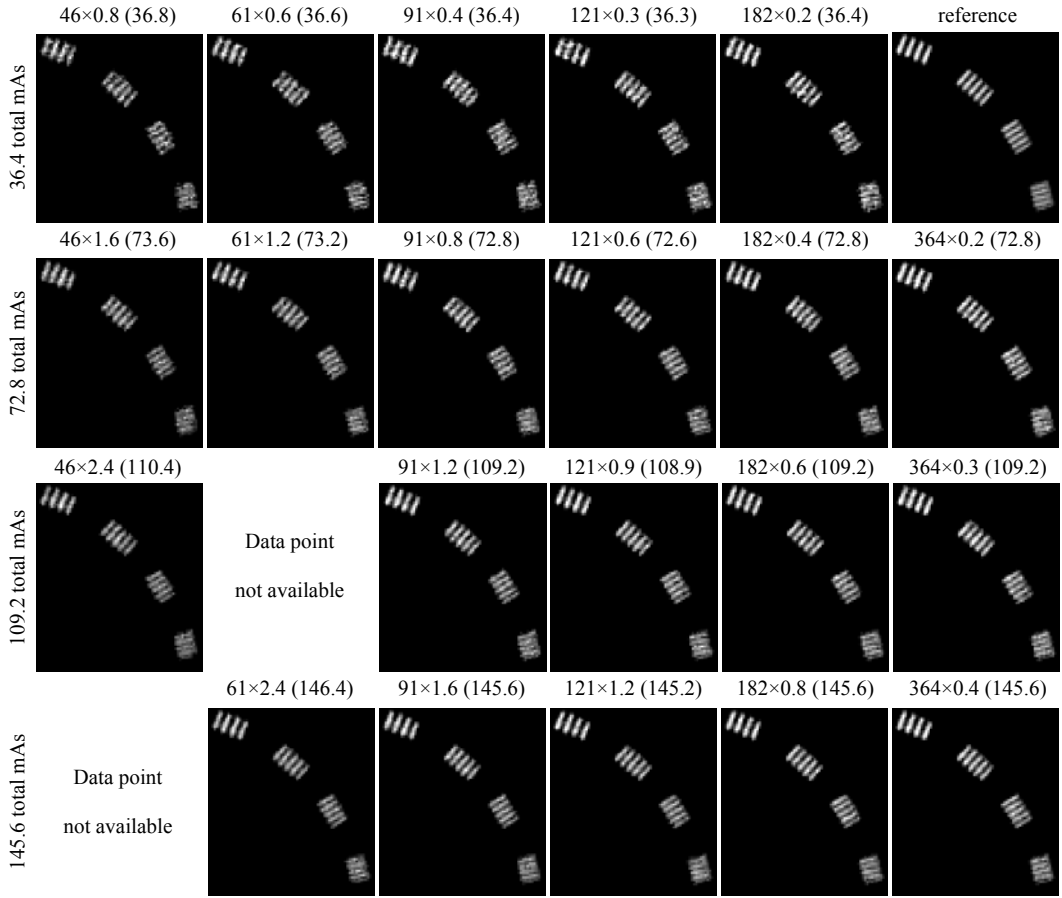

**Figure 11.** Reconstructed ROI images of the resolution slice with different scan protocols. Top to bottom: four iso-dose levels around 36.4, 72.8, 109.2 and 145.6 total mAs. Display grayscale: [0.03, 0.05] mm$^{-1}$.

*3.2.4 Towards extremely low dose: comparisons between different imaging tasks*

The reconstructed images acquired at extremely low dose levels are shown in figure 12 for different imaging tasks and compared with the reference images acquired at the highest dose level (873.6 total mAs). It clearly shows that neither 12.2 nor 18.2 total mAs is sufficient for the low-contrast or high-resolution objects. However, with even 12.2 total mAs we can still clearly see the high-contrast objects with large enough dimension. *e.g.*, 3 mm diameters for the small Teflon and air rob. This is consistent with the CNR results for the high-contrast objects shown in figure 10, *i.e.*, the visibility of high-contrast ROIs suffers the least from the dose reduction. Therefore, it may be quite feasible to use an extremely low dose level like 12.2 total mAs for certain IGRT tasks such as bony structure based patient positioning.

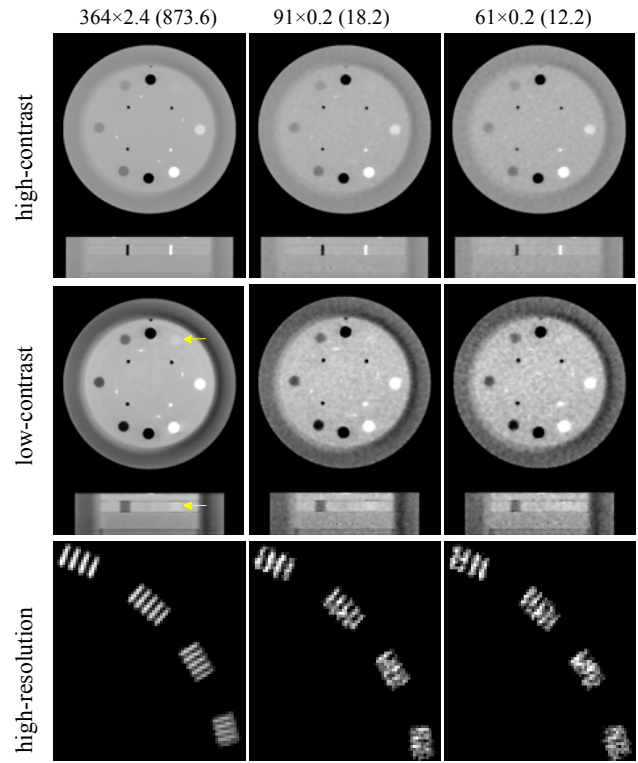

**Figure 12.** Top row: images with high-contrast objects with display grayscale [0.008, 0.032] mm$^{-1}$; middle
5    row: images with low-contrast objects (as indicated by the yellow arrows) with display grayscale [0.018, 0.025] mm$^{-1}$; bottom row: high-resolution ROIs with display grayscale [0.03, 0.05] mm$^{-1}$; left column: highest dose level used in the experiment (873.6 total mAs); middle and right columns: extremely low dose levels (18.2 and 12.2 total mAs, respectively).

## 4. Discussion

10        The results of this paper explicitly suggest that a scan protocol with a medium number of projections and a medium mAs/view level leads to the best image quality for a given low dose level. This phenomenon becomes more profound at lower dose levels. The underlying reason might be that this scan protocol provides an optimal balance between the view sampling requirement under the CS framework and the SNR demand in
15   each projection, which mitigates the under-sampling streaking artifacts and the excessive noise in the reconstructed images. In contrast of putting more focus on the low-contrast objects, another perspective might lead to a different observation. For example, focusing on the elimination of the streaking artifacts, Tang *et al* have suggested distributing the total dose into many view angles (Tang *et al.*, 2009). However, this suggestion is based
20   on the assumption that each view remains quantum noise limited, which may not hold for very low-dose cases in this work.

          The lowest possible dose that can still generate images of clinically acceptable quality is clearly task dependent. To ensure the visibility of the low-contrast objects, 72.8

total mAs seems sufficient (figure 5, 6 and 10), which is at least 50% of the dose used in the current low-dose protocol for head-neck cases (125kVp, 0.4mAs/view, 616 views over 360 degrees) (Murphy *et al.*, 2007). The dose level may be further reduced, *e.g.*, to 50 total mAs, under certain conditions, because the image quality observed for 36.4 total mAs seems acceptable too (figure 5). However, dose reduction below ~40 total mAs should be cautious since it may lead to dramatic image quality degradation (figure 3e and figure 4b). For the visibility of a high-contrast object with diameter $\geq$ 3 mm, the dose level can go down to 12.2 total mAs (figure 12).

As indicated from figures 5-8, the sparse view reconstruction is still the most challenging case for a given low dose level. At the same time, we can see that once the view numbers exceed a certain threshold, the streaking artifacts disappear and the image quality is improved dramatically. In the following text, we would like to present some intuitive interpretation on this observation. It has been reported that there exists redundancy among cone beam projections of neighboring view angles (Yan *et al.*, 2010). Therefore, it is reasonable to sample the object information by using a larger angular interval, namely, performing a sparse view reconstruction. As far as the minimum adequate view number is concerned, it differs from the classic object-independent view sampling requirement (Joseph and Schulz, 1980) based on Shannon-Nyquist theorem; instead, under the CS framework, it is supposed to be an issue highly relevant to the image resolution and the sparseness of the object. In this paper, 91 projections can effectively eliminate the under-sampling streaking artifacts for the CatPhan® 600. In (Bian *et al.*, 2010), 60 and 96 projections are needed for the simple cylinder and anthropomorphous phantoms. In (Song *et al.*, 2007), 95 retrospective gated projections provide sufficient quality for the cardiac imaging of one phase. In (Tang *et al.*, 2009), more than 100 projections are necessary to generate streaking-free images for the cadaver head. (Yan *et al.*, 2010) also reveals that redundancy within neighboring projections is almost squeezed out for a complex head phantom when the view number is decreased to 180 over 360 degree. We therefore empirically recommend 90-180 as the range for the project number that should be used for low dose protocols. Within this range, a larger view number should be used when the object is more complex or higher image resolution is needed. It will be interesting to study the minimum view number required for various resolution needs and different complexities of the scanned objects.

This work is intended to serve as a guideline for low dose CBCT imaging in IGRT. While some general observations, such as that there exists an optimal combination of mAs/view and view numbers at intermediate levels for a given total mAs value, are likely to hold regardless of disease sites or imaging systems, some quantitative conclusions may be dependent on disease sites and imaging systems and deserve further studies. For example, while the quantitative values obtained in this work may be applicable to head-and-neck cases on the Varian OBI systems, further studies are needed for the thorax/pelvis cases with the half-fan mode, and for other IGRT systems. Furthermore, the same methodology will be extended to more clinically realistic phantoms.

**5. Conclusions**

The conclusions of this work are as follows:
1) Under the CS-based reconstruction framework, image quality has little degradation over a large range of dose variation. Image quality degradation becomes evident when dose is decreased below 100 total mAs. An imaging dose lower than 40 total mAs leads to a dramatic image degradation, which should be used cautiously. Optimal low-dose CBCT scan protocols likely fall in dose range of 40-100 total mAs, depending on the specific IGRT applications.
2) Among the different scan protocols for a constant low-dose level, the super sparse-view reconstruction with projection number < 50 is the most challenging case, even with strong regularization. Better image quality can be acquired with the low mAs protocol.
3) The optimal scan protocol is the combination of a medium number of projections and a medium level of mAs/view. This is more evident when the dose is around 72.8 total mAs or below and when the ROI is a low-contrast or high-resolution object. Based on our results, the optimal number of projections is around 90 to 120.
4) The visibility of an object relies on its contrast and dimension, and hence the clinically acceptable lowest imaging dose level is task dependent. In our study, 72.8mAs is a safe dose level for visualizing the low-contrast objects, while 12.2 total mAs is sufficient for detecting high-contrast objects of diameter ~ 3 mm.

**Acknowledgements**

This work is supported in part by NIH (1R01CA154747-01), Varian Medical Systems through a Master Research Agreement, and the Thrasher Research Fund. The authors would like to thank the suggestions from Prof. Xuanin Mou about the IQA methods.